\documentclass[a4paper]{article}

\begin{document}

\title{\textbf{The Relativistic Quantum Stationary Hamilton
Jacobi Equation for Particle with Spin $1/2$. }}

\author{T.~Djama\thanks{E-mail:
{\tt djam\_touf@yahoo.fr}}}

\date {October 29, 2003}

\maketitle


\begin{abstract}
\noindent For one dimensional motions, we derive from the Dirac
Spinors Equation (DSE) the Quantum Stationary Hamilton-Jacobi
Equation for particles with spin $1/2$. Then, We give its
solution. We demonstrate that the $QSHJES_{1\over2}$ have two
explicit forms, which represent the two possible projection of
the Spin $1/2$.
\end{abstract}

\vskip\baselineskip



\newpage

\vskip\baselineskip \noindent
\textbf{1- Introduction}
\vskip\baselineskip
%
%

During about three years, a new deterministic approach of quantum
mechanics was arising from the works of Djama and Bouda
\cite{B-D1,B-D2,B-D3,Flo1}. It consist on the introduction of a
quantum Lagrangian with which and basing on the Quantum Stationary
Hamilton-Jacobi Equation (QSHJE), they derived the Quantum
Newton's Law in one dimension \cite{B-D1}. They also plot the
quantum trajectories for different potentials \cite{B-D2}. After,
Djama has generalize this formalism to the relativistic cases
\cite{Djama1,Djama2}. First, he took up the relativistic QSHJE
(RQSHJE) already introduced by Faraggi and Matone \cite{FM1}.

\begin{eqnarray}
{1 \over 2m_0}\left({\partial S_0 \over \partial x}\right)^2-
{\hbar^2 \over 4m_0}\left[{3 \over 2} \left({\partial S_0 \over
\partial x}\right)^{-2} \left({\partial^2 S_0 \over \partial
x^2}\right)^2-\right.
\hskip35mm&& \nonumber\\
\left.\left({\partial S_0 \over \partial x}\right)^{-1}
\left({\partial^3 S_0 \over \partial x^3}\right) \right]+ {1
\over 2m_0c^2}\left[m_0^2c^4 -(E-V)^2\right]=0\; ,
\end{eqnarray}
and introduced the relativistic quantum Lagrangian written as
\cite{Djama1}
\begin{equation}
L=-m_0c^2 \sqrt{1-f(x){\dot{x}^2 \over c^2}}-V(x)\; .
\end{equation}
From this Lagrangian, and using the least action principle, he
deduce the expression of the conjugate momentum in function of
the particle's velocity \cite{Djama1}
\begin{equation}
\dot{x}{\partial S_0 \over \partial x}=E-V(x)-{m_0^2c^4 \over 
E-V(x)}\; ,
\end{equation}
from which he derived the Relativistic Quantum Newton's Law

\begin{eqnarray}
\left[(E-V)^2-m_0^2c^4\right]^2+{\dot{x}^2 \over c^2}(E-V)^2
\left[(E-V)^2-m_0^2c^4\right]+{\hbar^2 \over 2} \left[{3 \over
2}\left({\ddot{x} \over \dot{x}}\right)^2-
{\dot{\ddot{x}} \over \dot{x}}\right] \cdot \hskip-10mm&& \nonumber\\
(E-V)^2-{\hbar^2 \over 2}\left(\ddot{x}{dV \over dx}+
\dot{x}^2{d^2V \over dx^2}\right) \left[{(E-V)^2+m_0^2c^4 \over
(E-V)^2-m_0^2c^4}\right] (E-V)^2- {3\hbar^2 \over 4}\cdot
\hskip-1mm&& \nonumber\\
\left(\dot{x}{dV \over dx}\right)^2 \left[{(E-V)^2+m_0^2c^4 \over
(E-V)^2-m_0^2c^4}\right]^2- \hbar^2\left(\dot{x}{dV \over
dx}\right)^2{m_0^2c^4 \over (E-V)^2-m_0^2c^4} =0\; .
\end{eqnarray}

\noindent In a last paper \cite{Djama2}, Djama had drawn the
trajectories of a relativistic spinless particles for many
potentials, and established the existence of nodes throw which
all possible trajectories pass, even the purely relativistic one
\cite{Djama2}. It is useful to stress that those results are found
in one dimension.

\noindent It is clear that the introduction of quantum and
relativistic quantum trajectories in the concept of a
deterministic approach of quantum and relativistic quantum
mechanics is a great step to elaborate a deterministic theory,
but it is not sufficient. The generalization to spinning particle
case and three dimension problems must be investigated.

\noindent The object of this paper is to investigate -in one
dimension- the Law of motion of particles with spin $1/2$. With
this aim, we derive in Sec. 2 the RQSHJE for Spin $1/2$
(RQSHJES$_{1/2}$) from the Dirac Spinors Equation (DSE) written in
one dimension. We demonstrate that the RQSHJES$_{1/2}$ is
composed by two components, each one correspond to the projection
of the Spin $1/2$ ($m_s=+1/2$ and $m_s=-1/2$). Then, in Sec. 3,
we propose a Double solution of the RQSHJES$_{1/2}$ corresponding
to $m_s=+1/2$ and $m_s=-1/2$. Finally, in Sec. 4, we discuss our
results.

\vskip\baselineskip \noindent \textbf{ 2- Quantum Stationary
Hamilton-Jacobi Equation for a particle with Spin $1/2$ }
\vskip\baselineskip

In the beginning, the quantum systems are described by the
Schr{\"o}dinger equation

\begin{equation}
-{h^2 \over 2m} {d^2 \psi \over dx^2}+ V(x)=E\; .
\end{equation}
But later, after the discovery of the spin by Goodsmith
\cite{CHPO} and Pauli \cite{CHPO}, and for studying the spinning
particle's behaviour, Dirac proposed his famous Spinors equation,
written with Pauli representation as \cite{unkn}
\begin{equation}
i\; \hbar c\; \vec{\alpha}\; .\;
\vec{\nabla}(\Psi)=\left[E-V(\vec{r})-\beta m_0c^2\right]
\end{equation}
where $\alpha_i, i=(x,y,z)$ and $\beta$ are the Dirac matrices
\begin{equation}
\alpha_i=\begin{pmatrix}
      { 0&\sigma_i  \cr
        \sigma_i&0  \cr}
 \end{pmatrix}; \ \
 \beta=\begin{pmatrix}
      { I&\ \ 0  \cr
        0&-I  \cr}
 \end{pmatrix}, \ \
\end{equation}
for which $\sigma_i$ are the Pauli matrices (Eq. (9)), and
$I=\begin{pmatrix}
      { 1&0  \cr
        0&1  \cr}
 \end{pmatrix}$ is the identity matrix.

The Dirac equation is written, in one dimension, in the Appendix
of Ref. \cite{Kudo} as
\begin{equation}
-i\; \hbar c \ \ \sigma_{x} \ \ {d \psi \over dx} \ =
(E-V(x)-\sigma_{z}\ m_{0}c^2) \ \psi
\end{equation}
where
\begin{equation}
\sigma_{x}=\begin{pmatrix}
      { 0&1  \cr
        1&0  \cr}
 \end{pmatrix}; \ \
 \sigma_{y}=\begin{pmatrix}
      { 0&-i  \cr
        i&\ 0  \cr}
 \end{pmatrix}; \ \
 \sigma_{z}=\begin{pmatrix}
      { 1&\ 0  \cr
        0&-1  \cr}
 \end{pmatrix}
\end{equation}
are the Pauli matrix. $\psi = \begin{pmatrix}
      { \theta  \cr
        \phi  \cr}
 \end{pmatrix}$ is a the matrix of the wave functions $\theta$ and
 $\phi$. Eq. (7) is a matrix equation of two components describing the
state of particles of half spin with energy $E$ moving in one
dimension under the action of the potential $V(x)$. It contains
two scalar equations which can be deduced after decomposing
\begin{equation}
-i\ \hbar \ c {d\phi \over dx}= \left(
E-V(x)-m_{0}c^2\right)\theta(x)\ ,
\end{equation}
\begin{equation}
-i\ \hbar \ c {d\theta \over dx}= \left(
E-V(x)+m_{0}c^2\right)\phi(x)\ .
\end{equation}
Taking the expression of $\theta$ from Eq. (10) and replacing it
into Eq. (11), then, taking the expression of $\phi$ from Eq. (11)
and replacing it into Eq. (10), we find

\newpage
\begin{equation}
\hbar^2 c^2 {d^2 \theta \over dx^2}+\hbar^2 c^2{ {dV / dx}\over
E-V+m_{0}c^2}{d \theta \over dx}
+\left[(E-V)^2-m_0^2c^4\right]\theta(x)=0\; ,
\end{equation}
\begin{equation}
\hbar^2 c^2 {d^2 \phi \over dx^2}+\hbar^2 c^2 { {dV / dx}\over
E-V-m_{0}c^2}{d \phi \over dx}
+\left[(E-V)^2-m_0^2c^4\right]\phi(x)=0\; .
\end{equation}
Now, the next step is to establish the QSHJES$_{1 \over 2}$. In
this order, let us write the two wave functions as
\begin{equation}
\theta(x)=A(x) \left(\alpha_{+} e^{ {i \over \hbar}S_0}+
\alpha_{-} e^{-{i \over \hbar}S_0}\right),
\end{equation}
\begin{equation}
\phi(x)=B(x) \left(\beta_{+} e^{ {i \over \hbar}Z_0}+ \beta_{-}
e^{-{i \over \hbar}Z_0}\right),
\end{equation}
where $A(x)$, $B(x)$, $S_0(x)$ and $Z_0(x)$ are real functions of
$x$. $\alpha_{+}$, $\alpha_{-}$, $\beta_{+}$ and $\beta_{-}$ are
real constants.

\noindent Replacing Eqs. (14) and (15) into Eqs.(12) and (13)
respectively and decomposing it, we find
\begin{equation}
{\hbar^2 c^2 \over A}{d^2A \over dx^2}-c^2{\left({dS_{0} \over dx
}\right)^2}+\hbar^2 c^2{dV/dx \over E-V+m_{0}c^2}{1 \over A}{dA
\over dx}+\left[(E-V)^2-m_0^2c^4\right]=0\; ,
\end{equation}
\begin{equation}
{\hbar^2 c^2 \over B}{d^2B \over dx^2}-c^2{\left({dZ_{0} \over dx
}\right)^2}+\hbar^2 c^2{dV/dx \over E-V-m_{0}c^2}{1 \over B}{dB
\over dx}+\left[(E-V)^2-m_0^2c^4\right]=0\; ,
\end{equation}
and
\begin{equation}
A\  {d^2 S_{0}\  \over dx^2}+2\ {dA \over dx}\ {d S_{0} \over dx}+
{dV/dx \over E-V+m_{0}c^2}\ A\ {dS_{0} \over dx}=0\; ,
\end{equation}
\begin{equation}
B\ {d^2 Z_{0} \over dx^2}+2\ {dB \over dx}{d Z_{0} \over dx}+
{dV/dx \over E-V-m_{0}c^2}\ B\ {dZ_{0} \over dx}=0\; .
\end{equation}
We can show easily that Eqs. (18) and(19) can be integrated to
give
\begin{equation}
A(x)=k_{1} (E-V(x)+m_{0}c^2)^{1 \over 2}\left(d S_{0} \over dx
\right)^{-{1 \over 2}}\; ,
\end{equation}
\begin{equation}
B(x)=k_{2} (E-V(x)-m_{0}c^2)^{1 \over 2}\left(d Z_{0} \over dx
\right)^{-{1 \over 2}}\; ,
\end{equation}
where $k_{1}$ and $k_{2}$ are two real constants. Replacing Eqs.
(20) and ( 21) into Eqs. (16) and (17), we get

\begin{eqnarray}
{1 \over 2m_0}\left({dS_0 \over dx }\right)^2- {\hbar^2 \over
4m_0} \{S_{0},x\}+ {\hbar^2 \over 2m_{0}}(E-V+m_{0}c^2)^{1 \over
2}\ .
\hskip25mm&& \nonumber\\
\ .{d^2 \over dx^2} \left[(E-V+m_{0}c^2)^{-{1 \over 2}}\right] +
{1 \over 2m_0c^2}\left[m_0^2c^4 -(E-V)^2\right]=0\; ,
\end{eqnarray}
\begin{eqnarray}
{1 \over 2m_0}\left({dZ_0 \over dx }\right)^2- {\hbar^2 \over
4m_0} \{Z_{0},x\}+ {\hbar^2 \over 2m_{0}}(E-V-m_{0}c^2)^{1 \over
2}\ .
\hskip25mm&& \nonumber\\
\ .{d^2 \over dx^2} \left[(E-V-m_{0}c^2)^{-{1 \over 2}}\right] +
{1 \over 2m_0c^2}\left[m_0^2c^4 -(E-V)^2\right]=0\; ,
\end{eqnarray}
where
$$
\{f(x),x\}=\left[{3 \over 2} \left({df \over dx }\right)^{-2}
\left({d^2 f \over dx^2}\right)^2-\left({df \over dx}\right)^{-1}
\left({d^3 f \over dx^3}\right) \right]
$$
represents the schwarzian derivative of $f(x)$ with respect to
$x$. Eqs. (22) and (23) represent the two Relativistic Quantum
Stationary Hamilton Jacobi Equations for Spinning particle $(s={1
\over 2})$ (QSHJES$_{1 \over 2}$. One of these equations
correspond to the projection $m_{s}=+{1 \over 2}$ of the spin,
when the other correspond to the projection $m_{s}=-{1 \over 2}$.
It follows that the reduced actions $S_{0}$ and $Z_{0}$ correspond
to the two projections of the spin.

\noindent We can connect the wave functions $\theta$ and $\phi$ to
the reduced actions as follows
\begin{equation}
\psi =
  \begin{pmatrix}
      { \theta  \cr
        \phi  \cr}
  \end{pmatrix}=
  \begin{pmatrix}
      { k_{1} (E-V(x)+m_{0}c^2)^{1 \over 2}\left(d S_{0} \over dx
\right)^{-{1 \over 2}}\left(\alpha_{+} e^{ {i \over \hbar}S_0}+
\alpha_{-} e^{-{i \over \hbar}S_0}\right) \cr
        k_{2} (E-V(x)-m_{0}c^2)^{1 \over 2}\left(d Z_{0} \over dx
\right)^{-{1 \over 2}} \left(\beta_{+} e^{ {i \over \hbar}Z_0}+
\beta_{-} e^{-{i \over \hbar}Z_0}\right) \cr}
  \end{pmatrix}
\end{equation}
Now, let us discuss Eqs. (22) and (23). First, remark that,
compared to the RQSHJE (Eq. (1)), Eq. (22) and (23) contains
additional terms
\begin{equation}
term_{1}={\hbar^2 \over 2m_{0}}(E-V+m_{0}c^2)^{1 \over 2}\ {d^2
\over dx^2} \left[(E-V+m_{0}c^2)^{-{1 \over 2}}\right]
\end{equation}
for Eq. (22), and
\begin{equation}
term_{1}={\hbar^2 \over 2m_{0}}(E-V-m_{0}c^2)^{1 \over 2}\ {d^2
\over dx^2} \left[(E-V-m_{0}c^2)^{-{1 \over 2}}\right]
\end{equation}
for Eq. (23). Both $term_{1}$ and $term_{2}$ are proportional to
$\hbar^2$ which means that at the classical limit ($\hbar \to 0$)
they vanish and Eqs. (22) and (23) reduce to the Classical
Stationary Hamilton-Jacobi Equation (CSHJE). Secondly, these two
terms vanish in the case of a constant potential. So, Eqs. (22)
and (23) reduce to the QSHJE. More than this, the motion of a
spinning particles are not affected by the constant potentials
but only by potentials with a not vanishing gradient with respect
to $x$.

\noindent By this, we deduce that $term_{1}$ and $term_{2}$ are
directly related to the spinning behaviour of quantum particles
in one dimension.

\vskip\baselineskip \noindent \textbf{ 3- The solutions of the
RQSHJES$_{1 \over 2}$ } \vskip\baselineskip

Now, let us investigate the form of the reduced actions $S_{0}$
and $Z_{0}$ solutions of the RQSHJES$_{1 \over 2}$ (Eqs. (22) and
(23)). We propose as solutions of Eqs. (22) and (23) the
following functions
\begin{equation}
S_0=\hbar \arctan\left(a\ {\theta_{1}(x) \over
\theta_{2}(x)}+b\right) \; ,
\end{equation}
\begin{equation}
Z_0=\hbar \arctan\left(d\ {\phi_{1}(x) \over \phi_{2}(x)}+e\right)
\; ,
\end{equation}
where $a, b, d$ and $e$ are real constants. $\theta_{1}$ and
$\theta_{2}$ are two real and independent solutions of Eq. (12).
$\phi_{1}$ and $\phi_{2}$ are two real independent solutions of
Eq. (13). Expressions (27) and (28) are analogous to the
expression of the reduced action, solution of the RQSHJE and the
QSHJE \cite{B-D1,Djama1,FM1,Flo2}.

\noindent Now, we demonstrate our proposition.
 First, remark that Eqs. (22) and (23) have an analogous form.
To write Eq. (23) we can replace $S_{0}$ by $Z_{0}$ and
$(E-V+m_{0}c^2)$ by $(E-V-m_{0}c^2)$. Eqs. (27) and (28) are also
analogous. For this reason, in what follows, we exhibit
demonstration only for $S_{0}$ and Eq. (22) (for $Z_{0}$ and Eq.
(23) the demonstration can be done with the same manner). In this
order, we introduce the function
\begin{equation}
\theta = a\ \theta_{1}+b \ \theta_{2}\; .
\end{equation}
As $\theta_{1}$ and $\theta_{2}$ are solutions of Eq. (12),
$\theta$ is also solution of it. $S_{0}$ takes the form
\begin{equation}
S_0=\hbar \arctan\left(a\ {\theta(x) \over \theta_{2}(x)}\right)
\; .
\end{equation}
The wronskian $W$ of $\theta$ and $\theta_{2}$ is deduced from
Eq. (12)
\begin{equation}
W=\ \theta \  {d\theta_{2} \over dx}-\ \theta_{2} \  {d\theta
\over dx}=\ \alpha \ (E-V(x)+m_{0}c^2)\; ,
\end{equation}
where $\alpha$ is a real constant. Using this expression of the
wronskian and replacing Eq. (30) into Eq. (22), after calculating
the schwarzian derivative of $S_{0}$ with respect to $x$, one find
\begin{eqnarray}
{\hbar^2 \over 2m_0}{{W^2-(\theta \  {d\theta_{2} / dx}-\
\theta_{2} \  {d\theta / dx})^2 }\over (\theta^2+\theta_{2}^2
)^2}-{\hbar^2 \over 2m_{0}}(E-V+m_{0}c^2)^{1 \over 2}\ .
\hskip5mm&&\nonumber\\
\ .{d^2 \over dx^2} \left[(E-V+m_{0}c^2)^{-{1 \over 2}}\right]+
{\hbar^2 \over 2m_{0}}(E-V+m_{0}c^2)^{1 \over 2}\ .
\hskip20mm&&\nonumber\\
\ . {d^2 \over dx^2} \left[(E-V+m_{0}c^2)^{-{1 \over 2}}\right]
-{\hbar^2 \over 2m_0} \left[ {dV/dx \over E-V+m_{0}c^2}\ . \right.
\hskip20mm&&\nonumber\\
\ . {(\theta\, {d\theta_{2} / dx}-\ \theta_{2} \ {d\theta / dx})
\over \theta^2+\theta_{2}^2 }+  \left. { (\theta \ {d^2\theta_{2}
/ dx^2}+\ \theta_{2} \ {d^2\theta / dx^2}) \over
\theta^2+\theta_{2}^2 } \right]
\hskip15mm&&\nonumber\\
+{1 \over 2m_0c^2}\left[m_0^2c^4-(E-V)^2\right]=0 \; ,
\hskip40mm
\end{eqnarray}
which reduces to
\begin{eqnarray}
\theta \ \left\{ \hbar^2 c^2 {d^2 \theta \over dx^2}+\hbar^2 c^2{
{dV / dx}\over E-V+m_{0}c^2}{d \theta \over dx}
+\left[(E-V)^2-m_0^2c^4\right]\theta \right\}
\hskip15mm&&\nonumber\\
+\theta_{2}\ \left\{ \hbar^2 c^2 {d^2 \theta_{2} \over
dx^2}+\hbar^2 c^2{ {dV / dx}\over E-V+m_{0}c^2}{d \theta_{2}
\over dx} +\left[(E-V)^2-m_0^2c^4\right]\theta_{2} \right\}=0 \; .
\end{eqnarray}
As $\theta$ and $\theta_{2}$ are two solutions of Eq. (12), Eq.
(33) is automatically satisfied. Then, expression given by Eq.
(27) is the solution of the RQSHJE$_{1 \over 2}$ given by Eq.
(22). We note that, with same manner, we can demonstrate that
expression given by Eq. (28) is the solution of the RQSHJE$_{1
\over 2}$ given by Eq. (23).

\newpage

\vskip\baselineskip \noindent \textbf{4- conclusion}
\vskip\baselineskip

In this paper, we have present two new results. The first one is
the establishment of the two Relativistic Quantum Stationary
Hamilton-Jacobi Equations for a half spinning particle ($s={1
\over 2}$)\ (QSHJES$_{1 \over 2}$). The second result is the
resolution of the RQSHJE$_{1 \over 2}$.

\noindent The RQSHJE$_{1 \over 2}$ is composed of two equations
(Eq. (22) and (23)). Each one represent a projection of the spin
($m_{s}={1 \over 2}$ and $m_{s}={-{1 \over 2}}$).

\noindent both RQSHJE$_{1 \over 2}$ reduce to the classical
Hamilton-Jacobi equation when we take the classical limit ($\hbar
\to 0 $).

\noindent for a free particle the spinning terms (Eqs. (25) and
(26)) vanish and both RQSHJE$_{1 \over 2}$ reduce to the RQSHJE.
So, for the free particle, we can not distinguish between two
projection of spin until it pass throw a non vanishing gradient
potential.

\noindent Another interesting remark is the non relativistic case.
the question is: Can we deduce, from the RQSHJE$_{1 \over 2}$, the
spinning behaviour of a particle for a purely quantum case? The
answer is: Yes. Let us review the RQSHJE$_{1 \over 2}$ (Eqs. (22)
and (23)). For the non relativistic limit, $T<<m_{0}c^2$ (where
$T=E-V(x)-m_{0}c^2$ is the kinetic energy of the particle), Eq.
(22) reduces to
\begin{equation}
{1 \over 2m_0}\left({dS_0 \over dx}\right)^2-{\hbar^2 \over
4m_0}\{S_{0},x\}+ V(x)-E'=0
\end{equation}
which is the ordinary QSHJE. $E'$ is the quantum energy without
rest energy $m_{0}c^2$.

\noindent In the other hand side, Eq. (23) reduces to
\begin{eqnarray}
{1 \over 2m_0}\left({dZ_0 \over dx}\right)^2-{\hbar^2 \over
4m_0}\{Z_{0},x\}+{\hbar^2 \over 2m_{0}}(E'-V)^{1 \over 2}\ .
\hskip15mm&&\nonumber\\
\ .{d^2 \over dx^2} \left[(E'-V)^{-{1 \over 2}}\right]+
V(x)-E'=0\; ,
\end{eqnarray}
which is different from the ordinary QSHJE. So, it is clear that
Eq. (35) describe one of the two projections of the spin. Thus
there is two QSHJES$_{1 \over 2}$ to describe the spinning
particle with $s={1 \over 2}$ in the purely quantum cases. This
point will be more investigated in a next papers.

\newpage
\vskip\baselineskip
\noindent \textbf{REFERENCES} 

\end{document}